\documentclass{aastex}
\usepackage{emulateapj5}
\usepackage{epsf}
\usepackage{psfig}
\usepackage{graphicx}
\usepackage{lscape}
\usepackage{natbib}
\usepackage{enumerate}

\shorttitle{}
\shortauthors{Rich et al.}

\begin{document}
\title{Near-IR Polarized Scattered Light Imagery of the DoAr 28 Transitional Disk}

\author{Evan A. Rich\altaffilmark{1},
John P. Wisniewski\altaffilmark{1},
Satoshi Mayama\altaffilmark{2},
Timothy D. Brandt\altaffilmark{3,31},
Jun Hashimoto\altaffilmark{1},
Tomoyuki Kudo\altaffilmark{4}, 
Nobuhiko Kusakabe\altaffilmark{5},
Catherine Espaillat\altaffilmark{6},
Lyu Abe\altaffilmark{7},
Eiji Akiyama\altaffilmark{5},
Wolfgang Brandner\altaffilmark{8},
Joseph C. Carson\altaffilmark{9},
Thayne Currie\altaffilmark{4},
Sebastian Egner\altaffilmark{4},
Markus Feldt\altaffilmark{8},
Kate Follette\altaffilmark{10}
Miwa Goto\altaffilmark{11},
Carol A. Grady\altaffilmark{12,13,14},
Olivier Guyon\altaffilmark{4},
Yutaka Hayano\altaffilmark{4},
 Masahiko Hayashi\altaffilmark{5}, 
Saeko S. Hayashi\altaffilmark{4},
Thomas Henning\altaffilmark{8},
Klaus W. Hodapp\altaffilmark{15},
Miki Ishii\altaffilmark{5},
Masanori Iye\altaffilmark{5}, 
Markus Janson\altaffilmark{16},
Ryo Kandori\altaffilmark{5},
Gillian R. Knapp\altaffilmark{17}, 
Masayuki Kuzuhara\altaffilmark{18},
Jungmi Kwon\altaffilmark{19},
Taro Matsuo\altaffilmark{20},
Michael W. McElwain\altaffilmark{12},
Shoken Miyama\altaffilmark{21},
Jun-Ichi Morino\altaffilmark{5},
Amaya Moro-Martin\altaffilmark{22,23},
Tetsuo Nishimura\altaffilmark{4},
Tae-Soo Pyo\altaffilmark{4},
Chunhua Qi\altaffilmark{24},
Eugene Serabyn\altaffilmark{25}, 
Takuya Suenaga\altaffilmark{5,26},
Hiroshi Suto\altaffilmark{5},
Ryuji Suzuki\altaffilmark{5},
Yasuhiro H. Takahashi\altaffilmark{5,19},
Michihiro Takami\altaffilmark{27},
Naruhisa Takato\altaffilmark{4},
Hiroshi Terada\altaffilmark{4},
Christian Thalmann\altaffilmark{28}, 
Daigo Tomono\altaffilmark{4},
Edwin L. Turner\altaffilmark{3},
Makoto Watanabe\altaffilmark{29},
Toru Yamada\altaffilmark{30},
Hideki Takami\altaffilmark{5}, 
Tomonori Usuda\altaffilmark{5},
Motohide Tamura\altaffilmark{5,19}}

\altaffiltext{1}{Homer L. Dodge Department of Physics, University of Oklahoma, Norman, OK 73071}
\altaffiltext{2}{The Center for the Promotion of Integrated Sciences, The Graduate University for Advanced Studies (SOKENDAI), Shonan International Village, Hayama-cho, Miura-gun, Kanagawa 240-0193, Japan}
\altaffiltext{3}{Astrophysics Department, Institute for Advanced Study, Princeton, NJ, USA}
\altaffiltext{4}{Subaru Telescope, National Astronomical Observatory of Japan, 650 North A‘ohoku Place, Hilo, HI 96720, USA}
\altaffiltext{5}{National Astronomical Observatory of Japan, 2-21-1, Osawa, Mitaka, Tokyo, 181-8588, Japan}
\altaffiltext{6}{Department of Astronomy, Boston University, 725 Commonwealth Avenue, Boston, MA 02215, USA}
\altaffiltext{7}{Laboratoire Lagrange (UMR 7293), Universite de Nice-Sophia Antipolis, CNRS, Observatoire de la Coted'azur, 28 avenue Valrose, 06108 Nice Cedex 2, France}
\altaffiltext{8}{Max Planck Institute for Astronomy, K¨onigstuhl 17, 69117 Heidelberg, Germany}
\altaffiltext{9}{Department of Physics and Astronomy, College of Charleston, 66 George St., Charleston, SC 29424, USA}
\altaffiltext{10}{Kavli Institute for Particle Astrophysics and Cosmology, Stanford University, 382 Via Pueblo Mall, Stanford, CA 94305}
\altaffiltext{11}{Universit¨ats-Sternwarte Mu¨nchen, Ludwig-Maximilians-Universit¨at, Scheinerstr. 1, D-81679 Mu¨nchen,Germany}
\altaffiltext{12}{Exoplanets and Stellar Astrophysics Laboratory, Code 667, Goddard Space Flight Center, Greenbelt, MD 20771, USA}
\altaffiltext{13}{Eureka Scientific, 2452 Delmer, Suite 100, Oakland CA96002, USA}
\altaffiltext{14}{Goddard Center for Astrobiology}
\altaffiltext{15}{Institute for Astronomy, University of Hawaii, 640 N. A‘ohoku Place, Hilo, HI 96720, USA}
\altaffiltext{16}{Department of Astronomy, Stockholm University, AlbaNova University Center, Stockholm SE-10691, Sweden}
\altaffiltext{17}{Department of Astrophysical Science, Princeton University, Peyton Hall, Ivy Lane, Princeton, NJ 08544, USA}
\altaffiltext{18}{Department of Earth and Planetary Sciences, Tokyo Institute of Technology, 2-12-1 Ookayama, Meguro-ku, Tokyo 152-8551, Japan}
\altaffiltext{19}{Department of Astronomy, The University of Tokyo, 7-3-1, Hongo, Bunkyo-ku, Tokyo, 113-0033, Japan}
\altaffiltext{20}{Department of Astronomy, Kyoto University, Kitashirakawa-Oiwake-cho, Sakyo-ku, Kyoto, Kyoto 606-8502, Japan}
\altaffiltext{21}{Hiroshima University, 1-3-2, Kagamiyama, Higashihiroshima, Hiroshima 739-8511, Japan}
\altaffiltext{22}{Space Telescope Science Institute, 3700 San Martin Dr., Baltimore, MD 21218, USA}
\altaffiltext{23}{Center for Astrophysical Sciences, Johns Hopkins University, Baltimore MD 21218, USA.}
\altaffiltext{24}{Harvard-Smithsonian Center for Astrophysics, 60 Garden Street, Cambridge, MA, 02138, USA}
\altaffiltext{25}{Jet Propulsion Laboratory, California Institute of Technology, Pasadena, CA, 91109, USA}
\altaffiltext{26}{Department of Astronomical Science, The Graduate University for Advanced Studies, 2-21-1, Osawa, Mitaka, Tokyo, 181-8588, Japan}
\altaffiltext{27}{Institute of Astronomy and Astrophysics, Academia Sinica, P.O. Box 23-141, Taipei 10617, Taiwan}
\altaffiltext{28}{Institute for Astronomy, ETH Zurich, Wolfgang-Pauli-Strasse 27, 8093 Zurich, Switzerland}
\altaffiltext{29}{Department of Cosmosciences, Hokkaido University, Kita-ku, Sapporo, Hokkaido 060-0810, Japan}
\altaffiltext{30}{Astronomical Institute, Tohoku University, Aoba-ku, Sendai, Miyagi 980-8578, Japan}
\altaffiltext{31}{NASA Sagan Fellow}

\begin{abstract}

We present the first spatially resolved polarized scattered light H-band detection of the DoAr 28 transitional disk.  Our two 
epochs of imagery detect the scattered light disk from our effective inner working angle of 0$\farcs$10 (13 AU) out to 0$\farcs$50 (65 AU).  This inner working angle is interior to the location of the system's gap inferred by previous studies using SED modeling (15 AU).
 We detected a candidate point source companion 1$\farcs$08 northwest of the system; however, our second epoch of imagery strongly 
suggests that this object is a background star.  We constructed a grid of Monte Carlo Radiative Transfer models of the system, 
and our best fit models utilize a modestly inclined (50$^{\circ}$), 0.01 $M_{\odot}$ disk that has a partially depleted inner gap from the dust sublimation radius out to $\sim$8 AU.  Subtracting this best fit, axi-symmetric model from our polarized intensity 
data reveals evidence for two small asymmetries in the disk, which could be attributable to variety of 
mechanisms.  

\end{abstract}


\section{Introduction}

Detailed observations of the spectral energy distributions (SEDs) of young stellar objects (YSOs) have revealed 
numerous systems with deficits of near-infrared flux compared to primordial YSOs.  These pre-transitional and 
transitional disk systems are interpreted as having inner gaps and holes in their disks \citep{1989ESOC...33..423S, 2007ApJ...664L.111E, 2007ApJ...670L.135E}.  Both infrared scattered light imagery \citep{2006ApJ...636L.153F, 2010ApJ...718L..87T, 2011ApJ...729L..17H, can13, ave14} and sub-millimeter observations \citep{2011ApJ...742L...5A, 2014A&A...562A..26B, van14} have 
confirmed this basic architecture for pre-transitional and transitional disk systems, and revealed additional sub-structure 
that could provide information about the mechanism responsible for clearing regions in these disks \citep{van14}.

Numerous mechanisms have been suggested to explain the clearing and partial clearing associated with pre-transitional and 
transitional disk systems, including grain-growth \citep{2005A&A...434..971D}, photo-evaporation \citep{2001MNRAS.328..485C},  disk instability \citep{2007A&A...463..775P}, and perturbation from planetary companions \citep{2011ApJ...729...47Z}.  These mechanisms 
can also influence other regions of these disks such as the inner walls and surface structures.  For example, grain growth 
can create a rounded gap edge \citep{2012A&A...544A..79B} whereas planets can inflate gap edges, creating a wall that can shadow parts of the outer disk or induce significant back scattering \citep{2012ApJ...749..153J, JangCondell2013}.  Grain growth should also induce very bright sub-millimeter features in systems exhibiting gaps in the near-infrared, although this is not seen in 
Oph IRS 48 or SAO 206462 \citep{2014A&A...562A..26B,Perez2014}.  Additional morphological features such as spiral arms have been 
detected in these disks, and could be attributable to stellar and sub-stellar companions or gravitational instabilities \citep{2003MNRAS.339..577B}. Planet-induced perturbations should co-rotate with the planet; hence, detecting the rotation of these 
structure could help to distinguish the origin of some disk structures \citep{2011ApJ...729L..17H, lom15}.  
Additional morphological features observed in transitional disks, such as dust traps \citep{2014A&A...562A..26B}, can also be used to constrain the mechanism responsible for sculpting the spatial distribution of gas and dust in transitional disks.  Clearly, a first step that is needed to assess the clearing mechanism is to fully constrain the spatial distribution of small and large dust grains, as well as the gas, in individual disk systems.

The spatial distribution of small dust grains in numerous transitional disks \citep{2012ApJ...758L..19H, 2012ApJ...748L..22M, 2013ApJ...762...48G, 2013ApJ...767...10F, 2012ApJ...753..153K, 2011ApJ...729L..17H, 2010ApJ...718L..87T, 2014ApJ...783...90T,2013ApJ...772..145T, can13, ave14, 2015ApJ...799...43H} have been recently parametrized utilizing several facilities, including large programs like the Strategic Exploration of Exoplanets and Disks with Subaru (SEEDS) high-contrast imaging survey \citep{tam09}.  These observations have provided scattered light confirmation 
of the gapped nature of transitional disks \citep{2010ApJ...718L..87T}, revealed the presence of spiral structures 
\citep{2012ApJ...748L..22M, 2013ApJ...762...48G,Currie2014}, detected likely non-axisymmetric inner disks \citep{2012ApJ...753..153K,2013ApJ...772..145T}, found differences in some cases about 
the distribution of large versus small dust grain 
populations \citep{2012ApJ...758L..19H,2012ApJ...750..161D}, and the presence of a planet within a gap \citep{Currie2014}.

DoAr 28 is a K5-type object in the $\rho$ Ophiuchi association located at a distance of $\sim$139 pc \citep{mam08}, that 
has been identified as a transitional disk based on its SED \citep{2010ApJS..188...75M,2013ApJ...769..149K}.  The system 
is observed to be actively accreting at a rate of $4*10^{-9}$ $\frac{M_{\sun}}{year}$ \citep{2014ApJ...787..153K}, and 
has a disk with an inferred outer gap radius of 15 AU \citep{2013ApJ...769..149K} from analysis of its SED. 

In this paper, we present the first scattered light detection of the DoAr 28 transitional disk, in the H-band.  In Section 
2, we will discuss the scattered light observations and data reduction. In Section 3, we present our analysis of the resolved disk, constrain the presence of co-moving point sources, and model these data using Monte Carlo models.  Finally, we 
discuss the implications of our results in the broader context of other resolved transitional disk systems in Section 4.

\section{Observations and Reductions}

DoAr 28 was observed in two epochs on 2012 July 9 (2012 epoch) and 2014 June 9 (2014 epoch) as part of the SEEDS survey using HiCIAO \citep{2008SPIE.7014E..19H} in the H-band.  The data were obtained in quad Polarized Differential Imaging (qPDI) mode at four wave-plate positions (0$^\circ$, 22.5$^\circ$, 45$^\circ$, 67.5$^\circ$) in Angular Differential Imaging (ADI) mode.  Each observational frame contains four sub-images with each sub-image having a field of view of 5$\farcs$0 by 5$\farcs$0, with a pixel scale of 9.5 mas pixel$^{-1}$ and a FWHM of 0$\farcs$136 for the 2012 epoch and 0$\farcs$101 for the 2014 epoch.  The total 
ADI field rotation achieved was 19.5$^\circ$ for the 2012 epoch data and was 21.3$^\circ$ for the 2014 epoch data.

\subsection{HiCIAO Data Reduction}
\label{subsec:HiCIAO_reduc}

The data reduction process we employed for extracting polarized intensity (PI) disk images utilized the double differencing reduction technique described in \citet{2011ApJ...729L..17H}. To briefly review, the four sub-images of each frame contain two ordinary 
and two extra-ordinary images, which can be summed and subtracted from their $90^\circ$  counterparts to create -Q, +Q, -U, and +U images.  The Q and U frames were then rotated into a common orientation, corrected for instrumental polarization, and summed to create final Q and U images. Note that as our 2012 epoch imagery were obtained under non-optimal conditions, we only 
utilized the best 64 of the 76 observed frames for our summed imagery. The final PI images are computed 
using $PI = \sqrt{Q^2 + U^2}$.  As previously noted by 
\citet{2012ApJ...758L..19H}, the PSF convolved by seeing is not perfectly corrected by the AO-188 system, which produces a residual polarized halo.  We computed an artificial halo following the procedure outlined in \citet{2012ApJ...758L..19H}, scaled this to the observed aperture polarization of the system (P = $0.869\% \pm 0.012\%$), and subtracted it from the PI image to create final PI imagery.  

We searched for point-source companions to DoAr 28 using the ACORNS-ADI software package \citep{2013ApJ...764..183B}. We treated each of the four sub-images noted above as its own angular differential imaging sequence (ADI, Marois et al. 2006) and reduced them using the LOCI algorithm (Lafreniere et al. 2007) with the standard reduction parameters from Brandt et al. (2013).  This yielded four residual images, each corrected for partial flux subtraction.  We then averaged these PSF-subtracted images to produce a single high-contrast image.  We computed the standard deviation in annuli on this combined image to produce a contrast map and searched for 5$\sigma$ companions.  We found one companion candidate as described in Section 3.2; follow-up data showed it to be an unrelated background star.

\subsection{SMA Observations and Reduction}

We observed DoAr~28 with the Submillimeter Array (SMA) on 2011 March 16, using the Compact Configuration with 
six of the 6 m diameter antennas at 230 GHz (1.3~mm) with a full correlator bandwidth of 2 GHz, for a total integration 
time of 63 minutes.
Calibration of the visibility phases and amplitudes was achieved with
observations of the quasar 3C~279, at intervals of about 20 minutes.
Observations of Titan provided the absolute scale for the flux density
calibration. The data were calibrated using the MIR software
package.\footnote{http://www.cfa.harvard.edu/$\sim$cqi/mircook.html}  We
detected DoAr~28 with a flux density of 68.6$\pm$1.9~mJy. The double sideband system temperatures were 110 to 170 K.

\subsection{APO Observations and Reduction}

DoAr~28 was observed using the ARC Echelle Spectrograph (ARCES; \citealt{wan03}) at the Apache Point Observatory (APO) 3.5m telescope on 
2014 June 19, yielding a R$\sim$31,500 spectrum covering the spectral range of $\sim$3,600-10,000\AA\ . The data 
were reduced using standard IRAF techniques.  We extracted the order containing H$\alpha$ and continuum normalized 
these data, enabling us to characterize the line strength discussed in Section \ref{subsec:Model}.

\section{Analysis}

\subsection{Scattered Light Imagery}
\label{subsec:Scattered}

\begin{figure}
\epsscale{.60}
\plotone{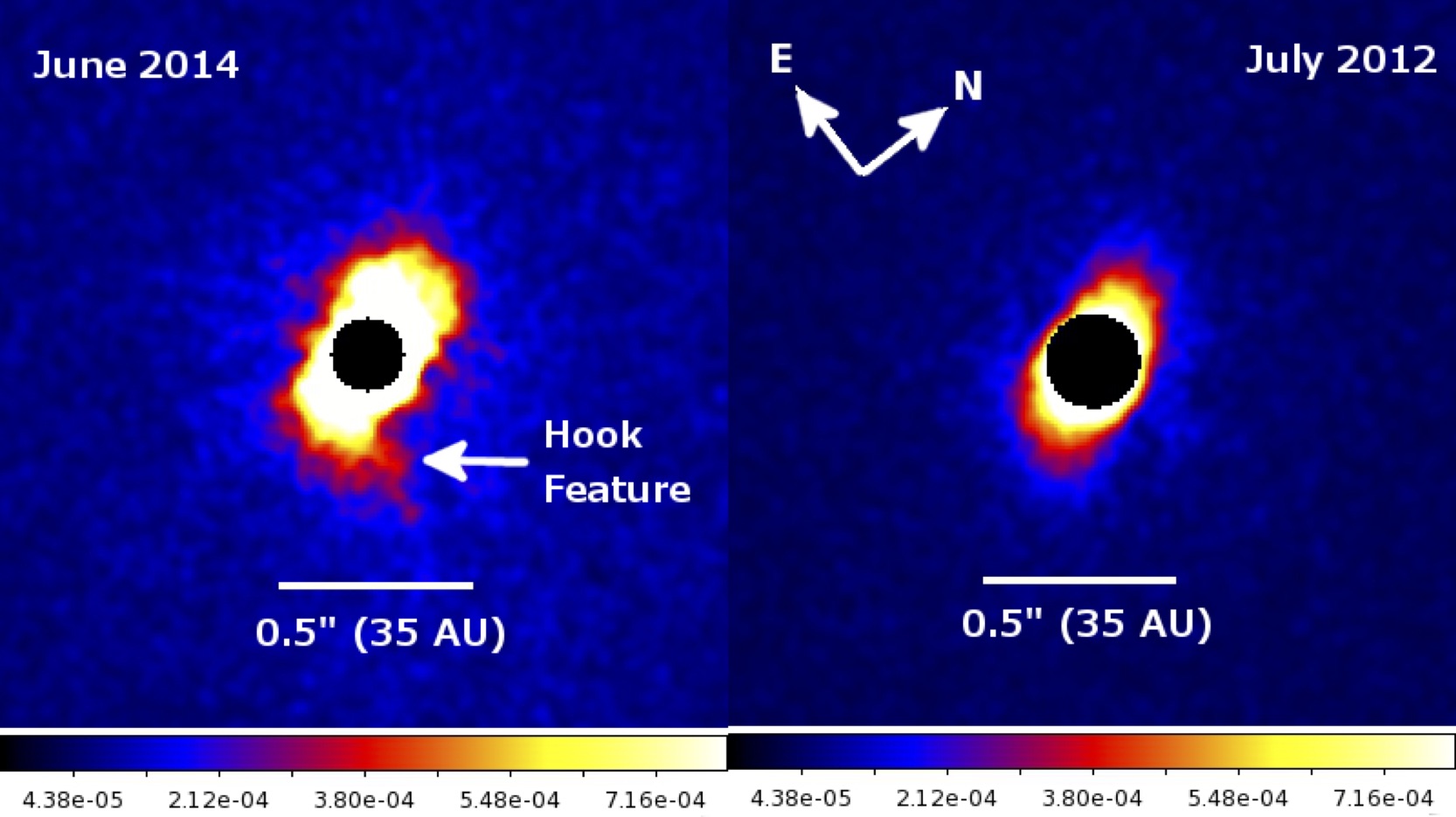}
\caption{The H-band polarized intensity (PI) image of DoAr~28 in 2014 June (left) and 2012 July (right) is shown over a field 
of view of 1$\farcs$95 x 1$\farcs$95 with the excess halo subtracted as described in section \ref{subsec:HiCIAO_reduc}.  We have applied a software mask over the location of the central star, with a size of 
0$\farcs$17 (24 AU) for the 2012 epoch data and 0$\farcs$10 (13 AU) for the 2014 epoch data, for asthetic 
purposes.  The imagery is plotted linearly in units of mJy, and was further smoothened with a 3-pixel Gaussian kernel.}\label{fig:PIimage}
\end{figure}

\begin{figure}
\epsscale{.30}
\plotone{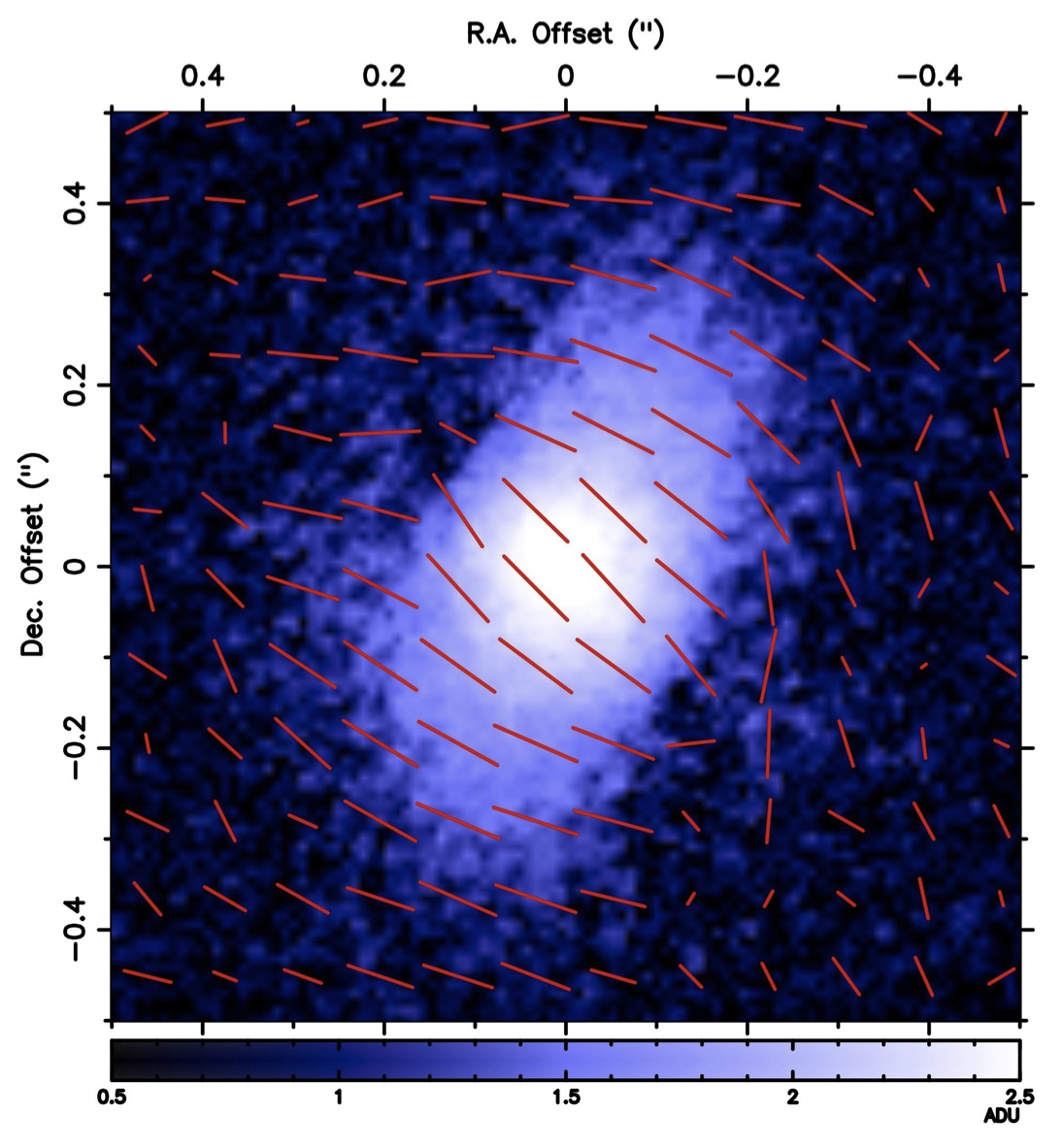}
\caption{A polarization vector map is overlaid on the PI image of DoAr~28 of epoch 2014 including halo subtraction. While the direction of the vectors represent the orientation of the observed polarization, the length of the vectors is a relative quantity and does not indicate the percent 
polarization present. The clear centrosymmetric behavior of the vectors confirms that we are detecting scattered light originating from the system's disk.}\label{fig:vector_map}
\end{figure}

Our two epochs of scattered light PI imagery of DoAr 28 are shown in Figure \ref{fig:PIimage}.  Significant scattered light  
around the central star is clearly seen in both epochs.  The direction and relative intensity of the polarization vectors 
derived from these data exhibit a clear centrosymmetric behavior around the central star (Figure \ref{fig:vector_map}), 
confirming that this signal arises from scattering off of circumstellar material.  These data exhibit less evidence of 
centro-symmetry about the minor axis, which is much less resolved than the major axis, suggesting that the level of our residual polarized halo correction might be incomplete.  Our primary analysis of these data will focus on the 
measured polarized intensity along the major axis of the system.  

\begin{figure}
\epsscale{.40}
\plotone{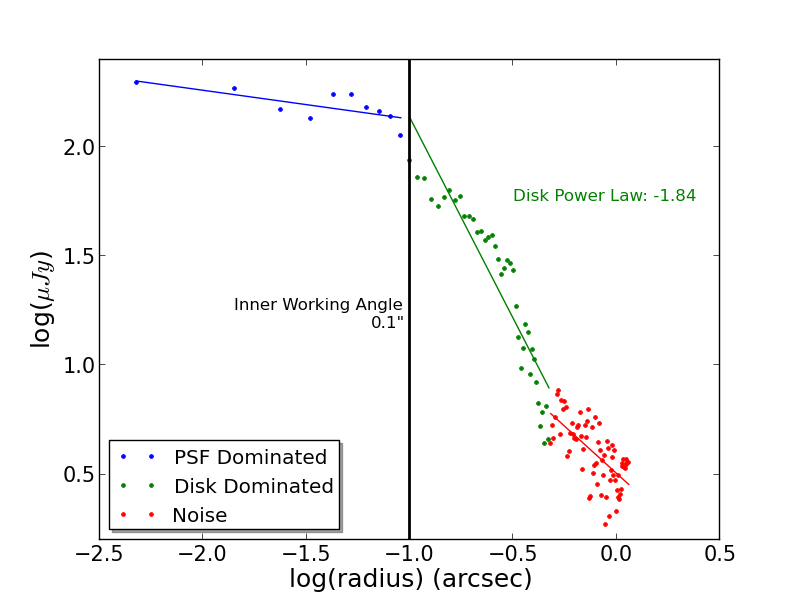}
\caption{The radial profile of 2014 epoch data along the north-side of the major axis illustrates clear evidence of two departures from 
a singular power law behavior.  One clear break in the power law happens at $\sim$0$\farcs$1 (the FHWM of the 2014 data), where the system transitions from 
being dominated by unsubtracted PSF residuals (blue) to disk dominated scattered light (green).  The second break occurs when the disk dominated scattered light (green) is overwhelmed by background noise (red).  The solid colored lines are best fit lines from linear regression fits to the disk power law shown on the figure. The black vertical line represents where the inner working angle is defined.} \label{fig:power_law}
\end{figure}

\begin{figure}
\epsscale{.40}
\plotone{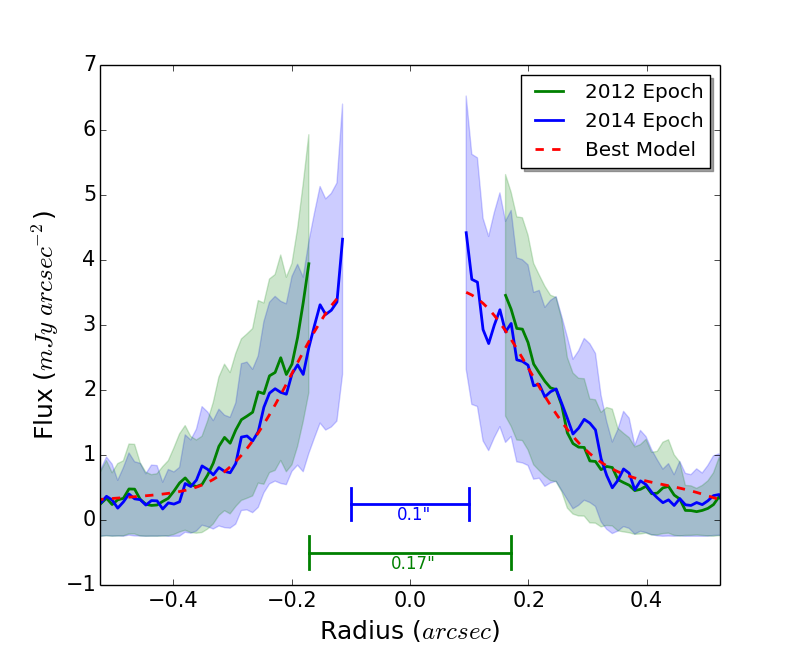}
\caption{The observed surface brightness along the major axis of the disk in our PI imagery at our 2012 (green) and 2014 epochs (blue), computed with a 4 pixel-wide average, is shown along with analogous surface brightness measurements from our MCRT model (red). The shaded colors represent a 1-sigma error bar with light green for 2012 epoch and light blue for 2014 epoch. The left side corresponds to the South end of the disk and the right side corresponds to the North side of the disk shown in Figure \ref{fig:PIimage}.  The central portions of these profiles have been removed, signifying the effective 
inner working angle of our data.}\label{fig:radial_profile}
\end{figure}

Since these data were not observed with a coronagraph, we defined the effective inner working angle as the 
location where the radial profile measured along the disk major axis (Figure \ref{fig:power_law}) exhibited a clear deviation from the disk dominated scattered light power law behavior seen in the outer disk.  Using this criterion, 
we determined the inner working angle to be 0$\farcs$17 (22 AU) for the 2012 epoch data and 
0$\farcs$10 (13 AU) for the 2014 epoch (Figure \ref{fig:radial_profile}).  The power law that characterizes the radial profile distribution of the scattered light flux along the disk major axis extends out to $\sim$0$\farcs$50 (65 AU), defining the outer 
edge of the scattered light disk, before becoming clearly dominated by noise.  The detected disk appears to be continuous, with no significant gaps or holes clearly visible.  The two epochs of imagery appear similar to 
each other, modulo potential differences arising from the factor of 1.4 better FWHM achieved in the 2014 epoch imagery. 
For the imagery with a slightly larger FWHM, it is conceivable that a small amount of disk flux could be spread out into the PSF halo, and be subtracted out in the process described in Section \ref{subsec:HiCIAO_reduc}.
Nevertheless, the overall surface brightness of the disk is the same at both epochs and the radial surface brightness power law 
measured along the major axis for both epochs is similar (-2.47 for 2012; -1.84 for 2014).  We do note the potential presence of a slight curl in the SW region of the 2014 epoch scattered light disk, and further discuss this feature in Section \ref{sec:Discussion}

\begin{figure}
\epsscale{.40}
\plotone{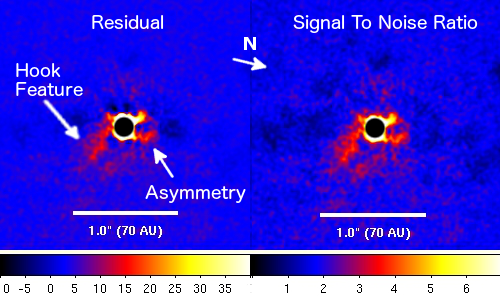}
\caption{We subtracted our best fit model from our 2014 epoch PI imagery, yielding the residual imagery shown on the 
left.  The right panel represents the signal to noise of the residual imagery.  
The image has been rotated such that the major axis is horizontal in the figure, with the northeast region of the disk on 
the right-hand side of the image.  Two potential asymmetries are observed, including a ``hook'' feature noted in Figure \ref{fig:PIimage} and a further asymmetry discussed in Section \ref{sec:Discussion}. The imagery is plotted linearly in units of ADU, and was further smoothened with a 3-pixel Gaussian kernel.}\label{fig:residual}
\end{figure}

We determine the inclination of the disk to be $\sim$50$^\circ$ (Table \ref{tbl:model}). This was found by subtracting a suite of
disk models for a range of inclination angles, described in Section \ref{subsec:Model}, and identifying the inclination which 
yielded the smallest residuals (Figure \ref{fig:residual}). Note that this residual image (Figure \ref{fig:residual}) exhibits clear deviations from axisymmetry, with a deficit of scattered light present along the northern side of the major axis.  We 
will be further discuss this non-axisymmetric structure in Section \ref{sec:Discussion}.

\subsection{Point Source Detections}
\label{sec:point_source}

We identified a candidate point source companion $1.\!\!''09$ northwest of DoAr 28, with an $H$-band contrast of 9.5 magnitudes, in our July 2012 data.  Our second epoch images from June 2014 indicate that this object is almost certainly a background star.  DoAr 28 was too faint for a proper motion from the {\it Tycho} satellite, but is part of to the same star forming region as $\rho$ Oph.  We assume that it shares $\rho$ Oph's proper motion of $(-5.5 \pm 0.9,\,-21.7 \pm 0.9)$ mas/yr in RA and Dec \citep{van07}.  At a distance of $\sim$140 pc, 1 mas/yr corresponds to $\sim$0.7 km\,s$^{-1}$, similar to the velocity dispersions of the Hyades, Pleiades, and TW Hya \citep{perryman1998, jones1970, makaov_fabricius2001}, and somewhat less than the $\sim$2--3 km\,s$^{-1}$ of Orion \citep{jones_walker1988, furesz_2008}.  Figure 6 shows that the companion candidate follows the expected background track assuming common proper motion with $\rho$ Oph.  If the point source were co-located with DoAr 28, their relative velocity would be at least 17 km\,s$^{-1}$, much too high for the system to be bound at its observed location.

Figure \ref{fig:contrast_curve} shows our sensitivity limits, assuming a distance of $\sim$140 pc and converting contrast to absolute magnitude, and neglecting extinction (estimated to be $\sim$0.4 magnitudes at H-band, using A(v) = 2.3 \citep{2010ApJS..188...75M}, R$_{v}$ = 3.1, and the \citet{ccm89} extinction relations), making these sensitivity estimates slightly optimistic.  It is more difficult to interpret these results as mass limits, given DoAr 28's extreme youth and uncertainties about the luminosities of very young planets \citep{2012ApJ...745..174S, 2011ASPC..448...91A, 2007ApJ...655..541M, 2008ApJ...683.1104F}.  Our SEEDS observations reach a limiting $H$-band absolute magnitude of $\sim$13.5 at a projected separation of 100 AU.
At an age of 5 Myr, this corresponds to $\sim$4 $M_{\rm Jup}$ in the BT-Settl models, but anywhere from $\sim$4 $M_{\rm Jup}$ up to the deuterium-burning limit of $\sim$13 $M_{\rm Jup}$ in the \cite{2012ApJ...745..174S} models (SB12), depending on the initial entropy.  We note that the high-mass end of this range requires a very cold start.  Assuming an initial entropy midway between the minimum and maximum values of the SB12 models, the SB12 mass limits are only slightly higher than the BT-Settl limits.

There are theoretical reasons arguing against the formation of substellar companions below $\sim$5 $M_{\rm Jup}$ by direct gravitational collapse \citep{1976MNRAS.176..367L, 2003MNRAS.339..577B, 2009MNRAS.392..590B}.  Our SEEDS imagery rule out such a companion, which would necessarily form hot and beyond $\sim$80 AU.  While a less massive core-accretion planet is unlikely to form at such a wide separation (DoAr 28's disk only extends to 70 AU), our data cannot rule out a Jovian planet scattered into a wide or unbound orbit.

\subsection{Radiative Transfer Modeling}
\label{subsec:Model}

We modeled the DoAr28 system with the HOCHUNK3D Monte Carlo Radiative Transfer (MCRT) code as described in Whitney et al. (\citeyear{2013ApJS..207...30W}). HOCHUNK3D is similar to other codes \citep{2009A&A...497..155M, 2003ApJ...596..603W, 2004A&A...421.1075D, 2006A&A...459..797P, 2011A&A...536A..79R,DAlessio2006} and has a long history of being used to constrain the dust distribution in protoplanetary systems. 
The latest version of HOCHUNK3D decouples the small and large dust grain distributions, allowing settling of dust to be incorporated. This implementation can be thought of as utilizing overlapping disks with different dust grain size distributions. 
Dust density distributions in HOCHUNK3D are adopted from Shakura \& Sunyaev (\citeyear{Shakura1973}) and are characterized by 
a radial power law ($\alpha$), and a vertical gaussian distribution ($\beta$), as given by equation \ref{eqn:density}, 
where $r$ is the radial component in cylindrical coordinates and $z$ is the height of the disk from the mid-plane. 
$h$ is the scale height defined in equation \ref{eqn:scale_height}, which is normalized at a defined radius of 100 AU. 
Deviations in the dust density distribution, such as gaps, spiral arms, and warped disks, can also be fully parameterized with HOCHUNK3D.

\begin{equation}
\label{eqn:density}
\rho \propto r^{- \alpha} \exp\left\{ \left[ \frac{z}{h}\right]^2\right\}
\end{equation}

\begin{equation}
\label{eqn:scale_height}
h \propto r^{-\beta}
\end{equation}

\citet{2013ApJS..207...30W} and references therein describe the full radiative transfer of the HOCHUNK3D code.  To 
briefly summarize, the code uses a Henyey-Greenstein scattering phase function and includes parameters for 
forward-scattering and albedo calculated from the adopted dust grain model.
The dust models we used are described below.  Temperature corrections utilized the Lucy method with a maximum number 
of six iterations \citep{lucy1999}.  We utilized 5*10$^{6}$ photons for our broad exploration of MCRT parameter space, and 5*10$^{7}$ photons for each of our runs where we compared both the observed SED and PI imagery against the models.

As is true with many MCRT codes, the large number of free parameters exceed the number of data points leading to model degeneracies. Since we lacked spatially resolved sub-millimeter observations that trace the radial distributions of large grains we assumed that the small and large grain disks had the same $\alpha$ and $\beta$ . We included a wall along the disk edge, as walls are thought to be common in transitional disks \citep{2005ApJ...630L.185C, 2007ApJ...664L.111E}. We also assumed that there was a negligible envelope, no warping of the disk, and that the disk was azimuthally symmetric. We adopted dust parameters from Wood et al. dust model 1 (\citeyear{2002ApJ...564..887W}) for our large dust grains, which are composed of amorphous carbon and silicon ranging in sizes up to 1 millimeter. The small dust grain model we adopted was the average galactic ISM model from Kim et al. (\citeyear{1994ApJ...422..164K}). We generated a grid of 270 models and identified a broad range of parameters consistent with the observed SED in Table \ref{tbl:model}.

Our model SED is consistent with the observed photometry of DoAr~28, as shown in Figure \ref{fig:SED}, with the new SMA data point helping to constrain the large dust grain disk. We explored a range of gap sizes and how these influenced the resultant SED and images, and demonstrate below that a gap size of $\sim$8 AU best represents our data.  We also found that the accretion rate and the gap density parameter had similar effects on the SED \ref{tbl:model}.
We adopted the accretion rate quoted by \citet{2014ApJ...787..153K}, $4.0*10^{-9}$ $\frac{M_{\sun}}{year}$, which 
led us assume a gap density parameter of 5*10$^{-5}$ (Table \ref{tbl:model}). Note however that our analysis of our new spectroscopic observations of DoAr 28 revealed a H-alpha 
  equivalent width ($30 \pm 0.1 \AA$) that was different than the 36 \AA\ reported in \citet{2014ApJ...787..153K}.  
  This suggests that the system likely exhibits a variable accretion rate. Thus, the upper and lower bound values of the gap density shown in Table \ref{tbl:model} are more representative of the system.

After using the observational SED of DoAr 28 to constrain our MCRT model parameters, we next used the observed surface brightness 
of our PI imagery to further constrain these parameters.  Specifically, we measured the surface brightness of our PI imagery using 
a 4-pixel wide aperture along the major axis of the disk, and compared this to the H-band PI surface brightness predicted by our 
models (Figure \ref{fig:radial_profile}).  This iterative process enabled us to arrive at our final adopted model parameters 
listed in Table \ref{tbl:model}, including the adoption of the gap size at $\sim$8 AU.
We used the model image subtracted from the PI images to find the inclination of 50$^{\circ}$. To search for potential deviations from axi-symmetry in our data, we subtracted our (axi-symmetric) model, scaled to the peak intensity of the observed PI disk, from our observations.  This process revealed evidence that the northern side of the disk exhibits a deficit of polarized flux near the inner working angle of our data as compared to the southern side of the disk (Figure \ref{fig:residual}).  We discuss the potential origin of this asymmetry in the discussion section.

\section{Discussion}
\label{sec:Discussion}

Our HiCIAO H-band multi-epoch observations of the DoAr~28 transitional disk clearly reveals evidence of a scattered light disk 
observed in polarized intensity, extending from our effective inner working angle of 13 AU (0$\farcs$10; 2014 epoch) to 
65 AU (0$\farcs$50).  We observe no gap in the disk in our PI imagery 
at the location suggested by previous SED modeling of the system (15 AU; 
\citealt{2013ApJ...769..149K}).  This suggests that either the small grain population is decoupled from the large grain 
population, and/or that the disk gap resides inside of our effective inner working angle.  
Our MCRT modeling of the disk, using constraints both from DoAr~28's SED and our H-band imagery, suggests
the disk has a gap extending from the dust sublimation radius out to $\sim$8 AU, that is only partially cleared of material.  
This gap size is smaller than that estimated from previous SED-only modeling efforts (15 AU; \citealt{2013ApJ...769..149K}).    
Other transitional disks observed in the H-band with HiCIAO as part of the SEEDS project, such as SAO 206462, MWC 758, and SR 21 \citep{2012ApJ...748L..22M, 2013ApJ...762...48G, 2013ApJ...767...10F}, 
also exhibit evidence of small dust grains in their gaps, similar to that inferred for DoAr~28.  Future multi-wavelength 
observations of the system that achieve a factor of $\sim$2 improvement in the effective inner working angle 
are needed to both test if the disk gap is truly as small as $\sim$8 AU, and to determine whether the radial distribution 
of small and large dust grains are decoupled, as is the case for systems like SR 21 \citep{2013ApJ...767...10F}.

The seemingly small gap size of DoAr~28 contrasts the disk gaps at larger orbital separations observed in other transitional 
disks, such as PDS 70 (70 AU; \citealt{2011ApJ...729L..17H}), SR 21 (36 AU; \citealt{2013ApJ...767...10F}), SAO 206462 
(46 AU \citealt{2012ApJ...748L..22M}), LkCa 15 (56 AU; \citealt{Thalmann2014}), and Oph IRS 48 (60 AU \citealt{2015ApJ...798..132F}). If this gap is caused by dynamical interactions between planetary bodies 
in the system and the disk, this could suggest that DoAr~28 has few such fully formed companions, perhaps owing to the system being in a more youthful state than other imaged systems.

DoAr~28 exhibits several indications of morphological features that deviate from simple axisymmetry at modest 
signal-to-noise levels (Figure \ref{fig:residual}).  For example, our 2014 
epoch imagery exhibits tentative evidence of asymmetry in the northern-side of the disk (Figure \ref{fig:residual}).  This type 
of asymmetry is commonly observed, and could be caused by a range of phenomenon ranging from companion interactions, a clumpy inner disk, and magneto-rotational instability.  A companion such as a planet could induce warping of the disk which would shadow the outer portions of the disk. If the observed asymmetric feature is caused by a companion, the feature should move in the disk at the same rotational speed of the companion, and not at the Keplarian speed \citep{2011ApJ...729L..17H}.  The asymmetry could 
also indicate the presence of a small azimuthal asymmetry in the inner disk.  Numerical simulations of magneto-rotations instabilities (MRI) also suggest that MRI-driven disk winds can perturb the disk \citep{2010ApJ...718.1289S}, which could also produce departures from axisymmetry. Additional epochs of spatially resolved imagery could help constrain whether the observed 
non-axisymmetric structure evolves with time, hence constrain the potential origin of this feature.

An additional, albeit more speculative indication of a non-axisymmetric feature is present along the southwest edge of the disk 
in our 2014 epoch data, namely the suggestive presence of a small hook-like feature (Figure \ref{fig:PIimage}).  
While the feature looks similar to spiral arms seen on in disks such as SAO 206462 and MWC 758 \citep{2012ApJ...748L..22M, 2013ApJ...762...48G}, there are no known T-Tauri transitional disks that exhibit spiral structures.  It has been suggested 
that the lower temperatures of T-Tauri disks (compared to their Herbig analogs) results in lower sound speeds, creating 
tighter spiral arms \citep{2012ApJ...748L..22M} that would not be resolved in our data.  Some studies have found that the dynamical temperature of CO gas is greater in T Tauri stars than in Herbig AeBe stars (\citeyear{2013A&A...559A..84M}), 
but this CO gas may not be well coupled with the small dust grain population.  We conclude that this hook, if 
it is real, is most likely to be a perturbation of the disk much like the asymmetric feature described above and not a 
spiral arm.

\section{Conclusion}

We have reported the first spatially resolved scattered light image of the DoAr~28 transitional disk in H-band. We detect the scattered light disk from 0$\farcs$10 (13 AU) out to 0$\farcs$50 (65 AU), which is slightly interior to the location of the system's gap inferred by another group's SED modeling (15 AU).  Although we detected a point source companion 
1$\farcs$1 northwest of the system, our second epoch of imagery of the system indicates this object is most likely a background
star.  Using the HOCHUNK3D Monte Carlo Radiative Transfer code, we have modeled both the observed SED and H-band PI imagery of 
the system.  Our best fit models utilize a modestly inclined (50$^{\circ}$), 0.01 M${\odot}$ disk that has a partially depleted inner gap from the dust sublimation radius out to $\sim$8 AU.  Subtracting this best fit, axi-symmetric model from our polarized intensity data reveals evidence for a small asymmetry in the northern-side of the disk, which could be attributable to variety of 
mechanisms.  We encourage future high spatial resolution sub-millimeter imagery of the system to better ascertain the location 
of the disk gap in the system, and to search for azimuthal and radial differences in distribution of small versus large dust 
grains that could be caused by recent planet formation in the system.   

We thank the referee for providing suggestions that improved the content and layout of this paper.  We acknowledge support from NSF-AST 1009203 (J.C.), 1008440 (C.G.), and 1009314 (E.R, J.W, J.H) and the NASA Origins of 
Solar System program under NNX13AK17G (J.W.), RTOP 12-OSS12-0045 (M.M.), and NNG13PB64P (C.G.).  This work is partly supported by a Grant-in-Aid for Science Research in a Priority Area from MEXT Japan and by the Mitsubishi Foundation. 
This work was performed [in part] under contract with the Jet Propulsion Laboratory (JPL) funded by NASA through the Sagan Fellowship Program executed by the NASA Exoplanet Science Institute.  The authors recognize and acknowledge the significant cultural role and reverence that the summit of Mauna Kea has always had within the indigenous Hawaiian community. We are most fortunate to have the opportunity to conduct observations from this mountain.

\clearpage

\begin{figure}
\epsscale{.40}
\plotone{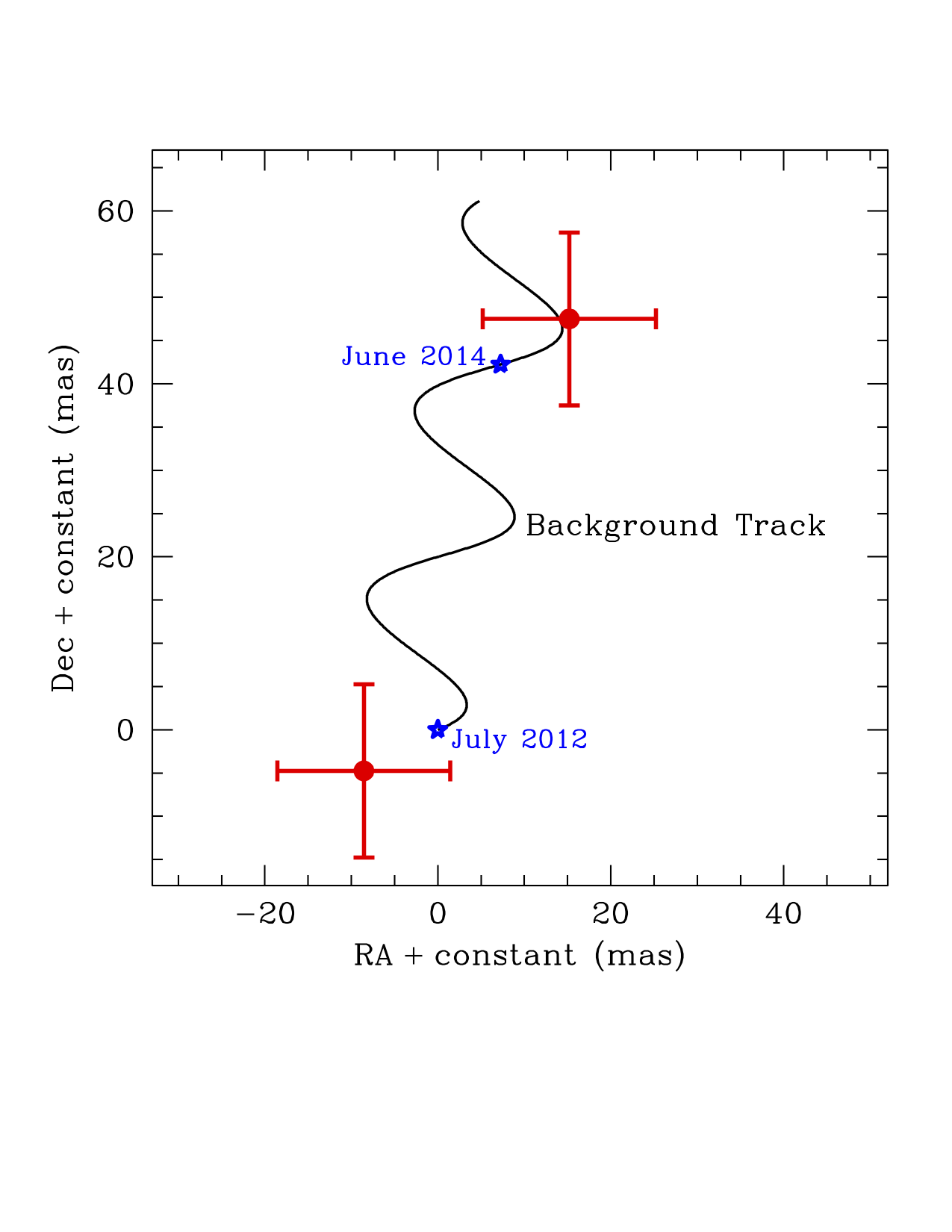}
\caption{The proper motion of DoAr 28 (blue star) and its candidate companion (red circle) in our
 two epochs of observations. DoAr 28 shares proper motion with $\rho$ Oph shown as the black sinusoidal line. The observed proper motion suggests that the companion is not co-located with DoAr 28, as 
 discussed in Section \ref{sec:point_source}.}
\label{fig:parallactic}
\end{figure}

\begin{figure}
\epsscale{.40}
\plotone{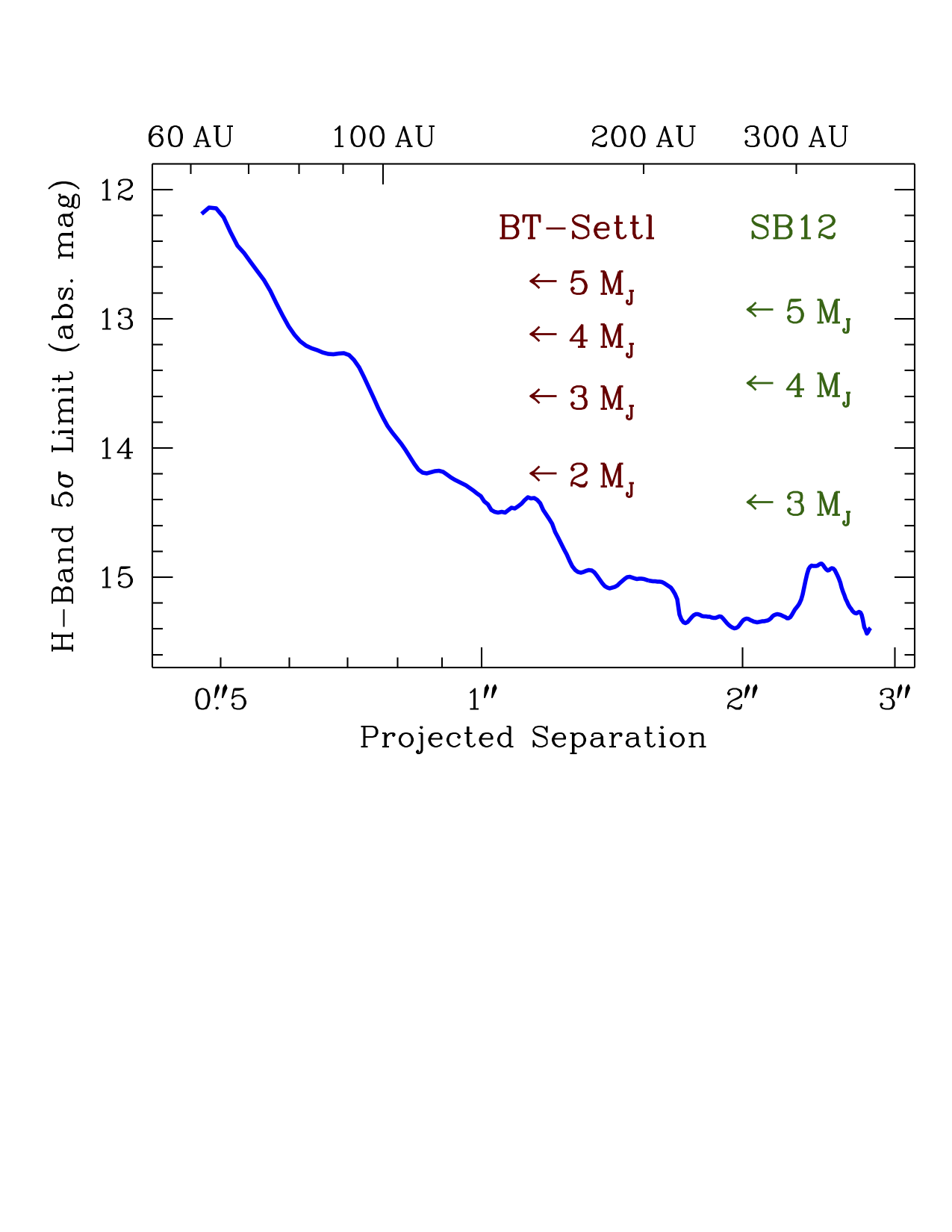}
\caption{The computed sensitivity limits of point source detections in our DoAr~28 imagery with a signal to noise of 5 sigma are given, along with associated planetary mass limits assuming both BT-Settl (burgundy) and SB12 (green) planet formation models. We assume an age of 5 Myr and SB12 ``warm start'' models midway between the minimum and maximum available initial entropies.
We do not detect evidence of any co-moving companions above these limits.}\label{fig:contrast_curve}
\end{figure}

\begin{figure}
\epsscale{.40}
\plotone{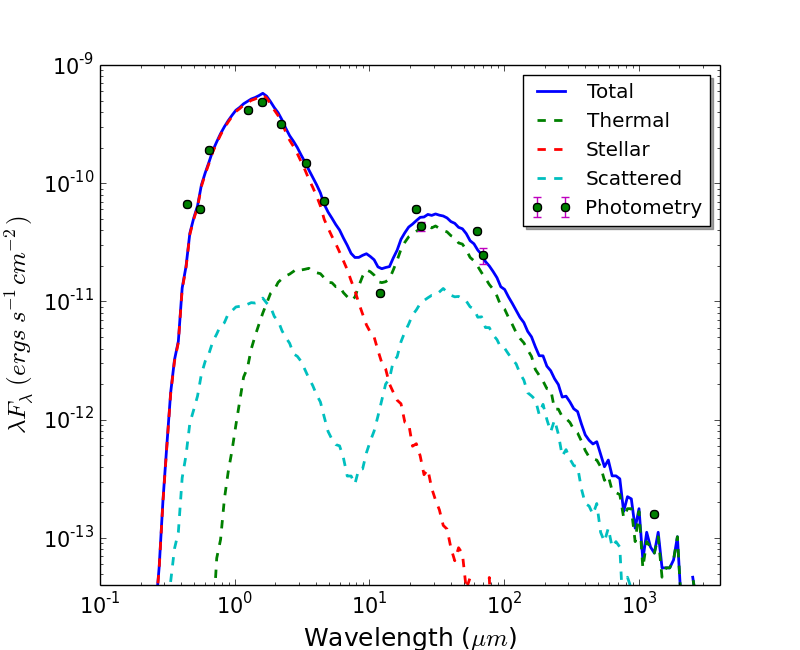}
\caption{The observed SED of DoAr~28, as compiled in Table \ref{tbl:SED}, is shown along with our best fit model, using 
the parameters compiled in Table \ref{tbl:model}. The solid line represents the total SED, whereas the dashed lines represent the different contributions to the SED, including thermal emission from the dust disk (thermal), the stellar photosphere (stellar), and scattered light from the disk (scattered).}
\label{fig:SED}
\end{figure}

\begin{table}
\begin{center}
\caption{Disk Model Parameters}
\label{tbl:model}
\begin{tabular}{crrrrrrrrrrr}
\\
\tableline
\tableline
Variable & Model & Low Bound & Upper Bound \\
\tableline
$T_{eff}$ (K) & 4375 & \nodata & \nodata \\
$R_{\sun}$ & 1.6 & 1.5 & 1.6 \\
Disk Mass  ($M_{\sun}$) & 0.01 & 0.005 & 0.05 \\
Max Disk Radius (AU) & 250 & 200 & 500 \\
Fraction of Mass & 0.8 & 0.7 & 0.99 \\
Inner Gap Radius (AU) & 0 & \nodata & \nodata \\
Outer Gap Radius (AU) & 8 & 7 & 9 \\
Wall height (AU) & 0.5 & 0.5 & 0.5 \\
Wall length (AU) & 0.5 & 0.5 & 1.1 \\
Zscale Disk 1 & 1.7 & \nodata & \nodata \\
Zscale Disk 2 & 1.7 & 1.7 & 1.9 \\
$\alpha$ & 2.0 & 1.9 & 2.1 \\
$\beta$ & 1.0 & 0.99 & 1.01 \\
Accretion ($\dot{M}_{\sun}$) & 4.0E-9 & \nodata & \nodata \\
Gap Density & 5E-6 & 5E-5 & 5E-7 \\
Inclination (i) & 50$^\circ$ & 40$^\circ$ & 65$^\circ$ \\
\tableline
\end{tabular}
\end{center}
\tablecomments{A summary of the lower and upper acceptable bounds of key HOCHUNK3D parameters that yield a fit consistent with
our observations, along with the ``best'' values that were adopted for our final model.  The ``fraction of mass'' parameter 
represents the fraction of the mass located in the large dust grain disk; the rest of the disk mass is located in the small dust grain disk.  The parameter ``zscale'' represents the scale height parameter. The ``gap density'' is the ratio of the dust density inside the gap relative to the density of dust at the inner edge of the disk. The "Zscale Disk" is the scale hight of disks 1 and 2 where disk 1 is large grain dust and disk 2 is small grain dust. Parameters $\alpha$ and $\beta$ describe the density profile of the disk defined in equation \ref{eqn:density}.}
\end{table}

\clearpage

\begin{table}
\begin{center}
\caption{Tabulated SED }\label{tbl:SED}
\begin{tabular}{crrrrrrrrrrr}
\\
\tableline\tableline
Wavelength ($\mu$m) & Flux (Jy) & Error (Jy) & Facility & Reference \\
\tableline

0.44 & 0.0098 & \nodata & Mt. Maidanak Obs. & \citep{gra07} \\
0.55 & 0.0112 & \nodata & Mt. Maidanak Obs. & \citep{gra07} \\
0.64 & 0.0411 & \nodata & USNO & \citep{mon98} \\
1.26 & 0.1759 & 0.0039 & 2MASS & \citep{2003yCat.2246....0C} \\
1.6   & 0.2606 & 0.0055 & 2MASS &\citep{2003yCat.2246....0C} \\
2.22 & 0.2379 & 0.0042 & 2MASS & \citep{2003yCat.2246....0C} \\
3.4 & 0.1670 & 0.0015 & WISE & \citep{2012yCat.2311....0C} \\ 
4.6 & 0.1092 & 0.00087 & WISE & \citep{2012yCat.2311....0C} \\
12.0 & 0.0471 & 0.00045 & WISE &\citep{2012yCat.2311....0C} \\
22.0 & 0.4450 & 0.0037 & WISE & \citep{2012yCat.2311....0C} \\
24.0 & 0.348 & 0.0322 & Spitzer/MIPS &\citep{2003PASP..115..965E} \\
62 & 0.829 & 0.013 & Herschel & \citep{2014ApJ...787..153K} \\
70.0 & 0.579 & 0.0891 & Spitzer/MIPS & \citep{2003PASP..115..965E} \\
1300 & 0.072 & 0.0019 & SMA & SMA Observation\\
\tableline
\end{tabular}
\tablecomments{The tabulated version of DoAr 28's photometry that was used to construct the SED shown in Figure \ref{fig:SED}.}
\end{center}
\end{table}

\end{document}